# Machine Learning a Phosphor's Excitation Band Position

Nakyung Lee[1,2,⊥], Małgorzata Sójka[1,2,⊥], Annie La[1,2], Syna Sharma[1,2], Seán Kavanagh[3], Docheon Ahn[4], David O. Scanlon[5], Jakoah Brgoch[1,2]

[1]Department of Chemistry, University of Houston, Houston, Texas 77204, USA
[2]Texas Center for Superconductivity, University of Houston, Houston, Texas 77204, USA
[3]Harvard University Center for the Environment, Cambridge, Massachusetts 02138, United States
[4]PLS-II Beamline Department, Pohang Accelerator Laboratory, POSTECH, Pohang 37673, Republic of Korea
[5]School of Chemistry, University of Birmingham, Birmingham, UK

[⊥]These authors contributed equally to this work

*jbrgoch@uh.edu (J. Brgoch)

*d.o.scanlon@bham.ac.uk (D. O. Scanlon)

**Abstract**

Creating superior lanthanide-activated inorganic phosphors is pivotal for advancing energy-efficient LED lighting and backlit flat panel displays. The most fundamental property these luminescent materials must possess is effective absorption/excitation by a blue InGaN LED for practical conversion into white light. The $5d_1$ excited state energy level of lanthanides, which determines the excitation peak position, is influenced by the inorganic host structure, including the local environment, crystal structure, and composition, making it challenging to predict in advance. This study introduces a new extreme gradient boosting machine learning method that quantitatively determines a phosphor's longest (lowest energy) excitation wavelength. We focus on the $Ce^{3+}$ $4f \rightarrow 5d$ transition due to its well-defined $5d_1$ energy level observed in excitation and diffuse reflectance spectra. The model was trained on experimental data for 357 $Ce^{3+}$ cation substitution sites sourced from literature and in-house measurements and ultimately experimentally validated through the successful synthesis of a novel, blue-excited, green-emitting phosphor: $Ca_2SrSc_6O_{12}:Ce^{3+}$. This compound's excitation under commercial blue LED wavelength aligned remarkably well with the model's predictions. These results highlight the transformative potential of data-driven approaches in expediting the discovery of blue-absorbing phosphors for next-generation LED lighting.

## 1. Introduction

Solid-state lighting, used today in applications ranging from general in-home and commercial illumination to high-resolution display technologies, was made possible by the development of blue-emitting indium gallium nitride (InGaN) LED chips.[1,2] Blue LEDs offer outstanding luminous efficacy, a small, durable architecture, a long usable lifetime, eco-friendly composition, and, most notably, serve as the starting point for producing energy-efficient white light. Indeed, converting this nearly monochromatic light source into a functional broad-spectrum white light is achieved by down-converting the LED's blue (~440 nm to ~470 nm) emission using one or more inorganic phosphors. The phosphors absorb and partially convert the blue light into usually yellow/green and red wavelengths, which in combination cover most of the visible spectrum, creating white light. Unfortunately, very few unique commercially produced phosphors are compatible with blue LEDs, with less than ten materials typically meeting the industry's

strict requirements.[3]

Designing new viable phosphors is an active and ongoing research effort that is an intrinsically multidimensional challenge, necessitating a deep understanding of phosphor chemistry.[4] Phosphors are inorganic host compounds, like oxides, nitrides, or halides, typically substituted with $Ce^{3+}$ or $Eu^{2+}$ that acts as the luminescence center or activator.[5–8] These two lanthanide ions are primarily used because they possess a broad emission with selection rule allowed $4f \leftrightarrow 5d$ electronic transitions. Depending on the interaction between the phosphor host and the activator ion, various optical responses can be achieved. Research has predominantly focused on understanding and predicting the factors controlling a phosphor's thermal quenching (temperature-dependent optical behavior) and photoluminescence quantum yield (PLQY), with the goal of identifying specific factors that support thermally robust photoluminescence and near-unity PLQYs.[9–13] Advances in high-throughput experimental and data-driven approaches have also improved the ability to produce desirable emission colors.[14–17] However, realizing phosphor's most critical property—its absorption of blue LED light—is still primarily guided by empirical understanding or costly first principles computational modeling.[18–20] Additionally, blue light absorption by lanthanide-substituted phosphors, especially $Ce^{3+}$, is relatively uncommon outside of the garnet family of materials. This has prompted researchers to instead consider UV (~365 nm) or violet (~400 nm) excited phosphors despite their limited potential for commercialization. The recent surge of phosphors published operating with these alternative LEDs underscores the need for new approaches that can spur the discovery of suitable blue-excited materials.

The position of the excitation band is set by crystal-chemical interactions between the inorganic phosphor host and the lanthanide ion. In the case of $Ce^{3+}$ and $Eu^{2+}$, these interactions cause the $5d$ orbitals to experience two well-known effects. The first concerns the host ligands (anions) stabilizing the centroid of the $5d$ orbital energy by withdrawing electron density and reducing electric repulsion.[21–25] This phenomenon, called the nephelauxetic effect or centroid shift, is dominated by the electronegativities and atomic polarizabilities of the coordinating anions, among other factors. The second effect, crystal field splitting, separates the nominally degenerate $5d$ orbitals based on the local coordination environment, such as coordination number, polyhedron volume, bond length, and coordination geometry (symmetry).[26–30] For example, lower coordination numbers with shorter bond lengths facilitate stronger crystal field splitting, as in octahedral or cubic environments, whereas larger sites like dodecahedral environments tend to have weaker crystal field splitting. The combined centroid shift and crystal field splitting generates a redshift ($D$), which represents the energy difference between the initial free ion ($Ce^{3+}$: 6.17 eV; 201 nm and $Eu^{2+}$: 4.22 eV; 294 nm) and the position of the lowest energy $5d$ orbital ($5d_1$ orbital) when substituted in the inorganic host.[31–34]

Considering the factors influencing the $5d_1$ energy level, and thus the excitation wavelength of the $Ce^{3+}$ or $Eu^{2+}$ phosphor ($4f \rightarrow 5d_1$), prior efforts have employed density functional theory (DFT) approaches to understand this transition.[35–37] These calculations are analogs to time-dependent DFT methods routinely employed for molecules, but they remain relatively uncommon in the solid state, primarily due to their high computational cost. In extended solids, particularly $Ce^{3+}$-doped phosphors, the energies and relative oscillator strengths of the $4f \rightarrow 5d$ transitions can be calculated using lanthanide-centered defect clusters within the host crystal structure. For example, prior work by our group and others have performed quantum chemical, wave-function-based complete active-space self-consistent-field (CASSCF) and second-order many-body perturbation theory (CASPT2) calculations on $Ce^{3+}$-centered embedded clusters.[38–43] These calculations have been used to understand how local coordination environments can influence a phosphor's absorption/excitation band position and reveal rare-earth substitution site preference.[37] The difficulty with these calculations is that they are mainly helpful in understanding a phosphor's excitation spectrum *after* their discovery. Moreover, their computational cost is high; large supercells and hybrid functionals are required to model the dilute lanthanide substitution. Consequently, it is impractical to use high-throughput calculations to uncover the structure-composition trends that guide lanthanide excitation band position, thereby limiting prediction or screening potential.

Here, the $5d_1$ excited state energy level position is effectively predicted through a new supervised machine learning model leveraging the extreme gradient boosting (XGBoost) algorithm. We constructed a dataset containing experimentally measured $5d_1$ excitation energies for $Ce^{3+}$ in 357 unique cation substitution sites across 337 different host materials, meticulously compiled from peer-reviewed studies and in-house measurements. $Ce^{3+}$ is the first lanthanide ion considered due to the straightforward experimental measurement of the redshift *D*, and therefore $5d_1$ energy, using optical spectroscopy. A feature set was engineered to reflect the dominant factors influencing the $4f{\rightarrow}5d_1$ optical transition by incorporating information on the local coordination environment, broader host crystal structure information, and chemical composition characteristics. Recursive feature elimination (RFE) and leave-one-group-out cross-validation (LOGO-CV) were employed to ensure model robustness and avoid over-fitting. The model's effectiveness for identifying blue-excited materials was then experimentally validated by screening for $Ce^{3+}$ substituted phosphors with excitation band positions that match standard InGaN LEDs. A promising candidate material, $Ca_2SrSc_6O_{12}$ substituted with $Ce^{3+}$, was identified, synthesized, and characterized to confirm the model's predictions, emphasizing the value of our new approach in guiding the discovery and design of future phosphors.

## 2. Experimental Methods

### 2.1 Prediction Model Construction

The model to predict $5d_1$ required data collection and feature engineering, as discussed below. Target values were collected from peer-reviewed publications regardless of the energy scale and then converted into eV for model training. Some of the most important features involved our group's previously reported machine learning prediction models for relative permittivity ($\varepsilon_r$, dielectric constant) and centroid shift ($\varepsilon_c$). These models were both updated and revised to get the most reliable input features for this work.[44] The relative permittivity model utilized a training data set of 2,254 values extracted from the Materials Project and cross-referenced with Pearson's Crystal Database to ensure only experimentally reported compounds were used in the training set. The range of $\varepsilon_r$ value was further limited to between 0 and 8, reflecting the typical relative permittivity range for most inorganic host structures. XGBoost was employed for the $\varepsilon_r$ model, and the model was validated through leave-one-out-cross-validation (LOO-CV)[45,46], achieving an $R^2$ of 88.7% (**Figure S1a**), comparable with the previously published model. Subsequently, the centroid shift model was updated using the revised predicted relative permittivity values, and $\varepsilon_c$ was similarly optimized with XGBoost, yielding a LOO-CV $R^2$ of 90.1% (**Figure S1b**).

The main model in this work, predicting the $5d_1$ energy, employed the tree-based XGBoost algorithm due to its demonstrated effectiveness in handling small and noisy datasets while mitigating overfitting. This was crucial considering the relatively modest size of the training set. Hyperparameter optimization was conducted with leave-one-group-out-cross-validation. It is necessary to group targets by composition because select targets with polymorphs or multiple substitution sites in a single phosphor host typically have very similar feature sets. LOGO-CV can prevent potential data leakage from this shared information. The optimization process encompassed 11 hyperparameters of the XGBoost model, including tree boosters that regulate individual tree models (*e.g.*, learning rate, max depth, and subsample), overfitting regularizations (*e.g.*, L1 and L2 regularization), and the base score parameter. The models were scored using mean absolute error (MAE) to quantify average model performance and assess actual deviations in predicted values, while recursive feature elimination was implemented to address multicollinearity and further limit overfitting.[47] After RFE, the feature set was reduced from 124 initial features to 44 scientifically relevant features (**Table S1**). The XGBoost model was finally retrained using this refined feature set, and hyperparameters were re-optimized with analysis by LOGO-CV. All model construction and implementation were done within the Python ecosystem, leveraging Scikit-learn libraries.[48]

## 2.2 Predicted Phosphor Synthesis and Characteristic Measurements

$Ca_2SrSc_6O_{12}$:$Ce^{3+}$ selected for experimental validation was synthesized by solid-state reaction starting from $CaCO_3$ (Alfa Aesar, 99.0%), $SrCO_3$ (Alfa Aesar, 99%), $Sc_2O_3$ (Thermoscientific, 99.99%), and $CeO_2$ (Sigma Aldrich, 99.995%). Each component was weighed in the appropriate stoichiometric ratio, with 1 mol% of the lanthanide activator added. The mixed starting reagents were ground in an agate mortar and pestle using hexane as a wetting medium and then further milled for 30 min in a high-energy ball mill (Spex 800 M Mixer/Mill). The mixture was pressed into a 6 mm diameter pellet and placed on a bed of sacrificial powder in an alumina crucible. The pellet was first heated to 1100°C for 10 h with a heating and cooling rate of 3°C/min under flowing 5% $H_2$/ 95% $N_2$ gas. The product was then ground in hexane, pelletized again, and heated a second time to 1300°C for 8 h with a heating and cooling rate of 3°C/min under flowing 5% $H_2$/ 95% $N_2$ gas.

The product was characterized using powder X-ray diffraction on an X'Pert PANalytical Empyrean 3 equipped with Cu Kα radiation (λ = 1.54056 Å). Additionally, high-resolution synchrotron powder X-ray diffractograms were collected at 9B beamline of PLS-II. Rietveld refinements were performed using the General Structural Analysis System II (GSAS-II) software.[49] The background was described using a Chebyshev-1 function, and the peak shapes were modeled using a pseudo-Voigt function. Photoluminescence measurements involved mixing the polycrystalline products in an optically transparent silicon epoxy (United Adhesives Inc., OP 4036) and depositing the combination onto a quartz slide (Chemglass). Photoluminescent excitation and emission and temperature-dependent luminescence measurements were obtained using a PTI fluorescence spectrophotometer with a 75 W xenon arc lamp for excitation. A Janis cryostat (VPF-100) was employed to establish a temperature-controlled environment from 80 K to 640 K. The photoluminescent quantum yield was determined following the method of de Mello *et al.*[50] using a Spectralon-coated integrating sphere (150 mm diameter, Labsphere). Photoluminescent lifetimes were measured using a 455 nm NanoLED equipped with the Horiba DeltaFlex Lifetime.

## 3. Results and Discussion

### 3.1 $5d_1$ Data Extraction and Feature Engineering

The construction of the machine learning model to predict the $5d_1$ excitation energy commenced with the compilation of 357 experimentally measured $5d_1$ energy levels, also reported as redshift values (*D*), from the literature as well as data collected in-house.[34] These data represent 337 individual host structures, including seven crystal polymorphs and compounds with multiple possible crystallographically independent cation substitution sites in a single host crystal structure. Diffuse reflectance and photoluminescence excitation spectra were both used to obtain reliable $5d_1$ target values. The distribution of collected target values shows that the training set continuously covers the entire excitation wavelength range from UV to blue/green (**Figure 1a**). Notably, most of the population is concentrated in the UV range, with only ~4% of $Ce^{3+}$ excitation target values falling in the typical blue LED emission range (440 nm - 470 nm), emphasizing the challenge of discovering blue-excited $Ce^{3+}$ phosphors. The excitation wavelengths can be analyzed based on their meta-data to understand the physical and chemical properties that dictate excitation wavelength. The relationship between $5d_1$ values and the coordination of the polyhedron occupied by $Ce^{3+}$, plotted in **Figure 1b**, highlights that higher $Ce^{3+}$ coordination number (>8) rarely generates excitation in the blue range due to smaller crystal field splitting energies (and thus lower $D$). A correlation between $5d_1$ values and the radius difference between $Ce^{3+}$ and the substituted cation further revealed that smaller sites tend to generate blue excitation due to stronger crystal field splitting (**Figure 1c**). When categorized by the host structure anion type, the target values also effectively demonstrate the trend of the nephelauxetic series[21,25] (**Figure 1d**), emphasizing the importance of compositional information such as electronegativity, electron affinity, and polarizability. This basic materials informatics analysis is already helpful for conceptualizing phosphor design and can also guide feature set engineering.

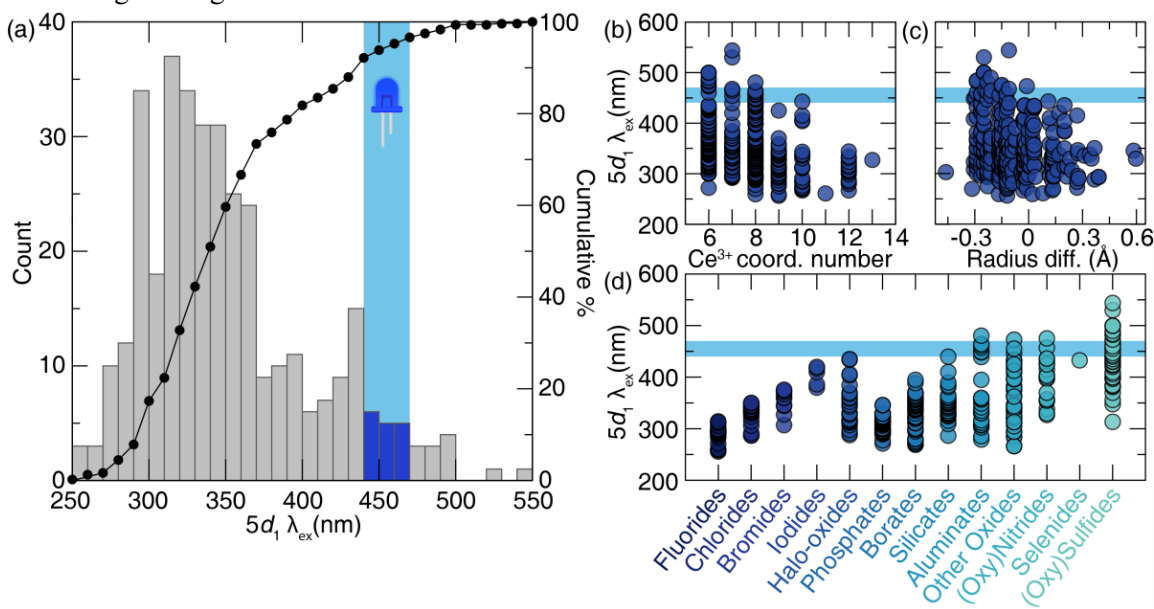

**Figure 1.** Target value distribution of the $Ce^{3+}$-substituted phosphors collected in the training set grouped by (a) excitation wavelength, (b) $Ce^{3+}$ coordination number of the crystallographically independent substitution sites, (c) ionic radius difference between the substituted cation and $Ce^{3+}$, and (d) the host anion type. The blue-shaded region highlights conventional emission wavelengths of blue-emitting InGaN LED chips.

An optimal set of relevant features was subsequently developed to create a quantitative machine-learning model. An initial 124 features were assembled, encompassing characteristics from the local environment of the cation substitution site (13 features), host crystal structure information (14 features), and host composition (95 features), with the complete set listed in **Table S1**. The local cation site features are necessary to estimate crystal field splitting for the $Ce^{3+}$ $5d$ orbitals. These features include the polyhedron volume, coordination number, and local symmetry. Features were also incorporated to describe the host crystal structure, which is essential for distinguishing polymorphic phosphors. These crystal structure attributes include space group number, volume of unit cell, volume per atom, volume per Z, and unit cell parameters, among other features. The cation substitution site and crystal structure information were extracted from the Materials Project database or experimental crystallographic information files from the Inorganic Crystal Structure Database (ICSD) and Pearson's Crystal Database (PCD).[51–53] In cases where the experimentally reported host crystal structure contains statistical site disorder, the *OrderDisorderedStructureTransformation* module from the Pymatgen library was used to predict a low-energy ordered arrangement from which the features were extracted.[54] The compositional features were finally developed by relating elemental properties to the stoichiometric composition. Each elemental feature (*e.g.*, polarizability, electron affinity, etc.) was expanded into five variables (maximum, minimum, average, difference, and standard deviation) to capture multidimensional relationships between features and target values. Two physics-based features were also added: relative permittivity ($\varepsilon_r$, dielectric constant) and centroid shift ($\varepsilon_c$) because they are understood to impact the nephelauxetic effect and, therefore, the $5d_1$ position.[44] These values depend explicitly on the host material and necessitated updated versions of our group's previously published supervised machine-learning regression models for these properties.

### 3.2 Machine Learning the $5d_1$ Energy Level

The $5d_1$ energy level prediction model was first constructed using target values from 357 unique cation sites. XGBoost was selected for its effective overfitting regulation methods since the training set contains only a few hundred target values. During model training, LOGO-CV was implemented to ensure appropriate inclusion of the training data while mitigating potential biases associated with *n*-fold cross-validation. The initial model with all 124 features showed an MAE of ±0.159 eV (RMSE = 0.160 eV) and an $R^2$ of 84.3 % (**Figure S2**). It is worth noting that this ±0.16 eV MAE manifests differently across the wavelength scale due to the non-linear relationship between energy and wavelength. For instance, the MAE when $5d_1$ is between 3.50 eV and 3.65 eV corresponds to a wavelength difference of only approximately 15 nm (from ~354 nm to ~339 nm), whereas at lower energy, the same energy gap can result in a more substantial wavelength difference, such as a 27 nm gap between 2.60 eV (~477 nm) and 2.75 eV (~450 nm).

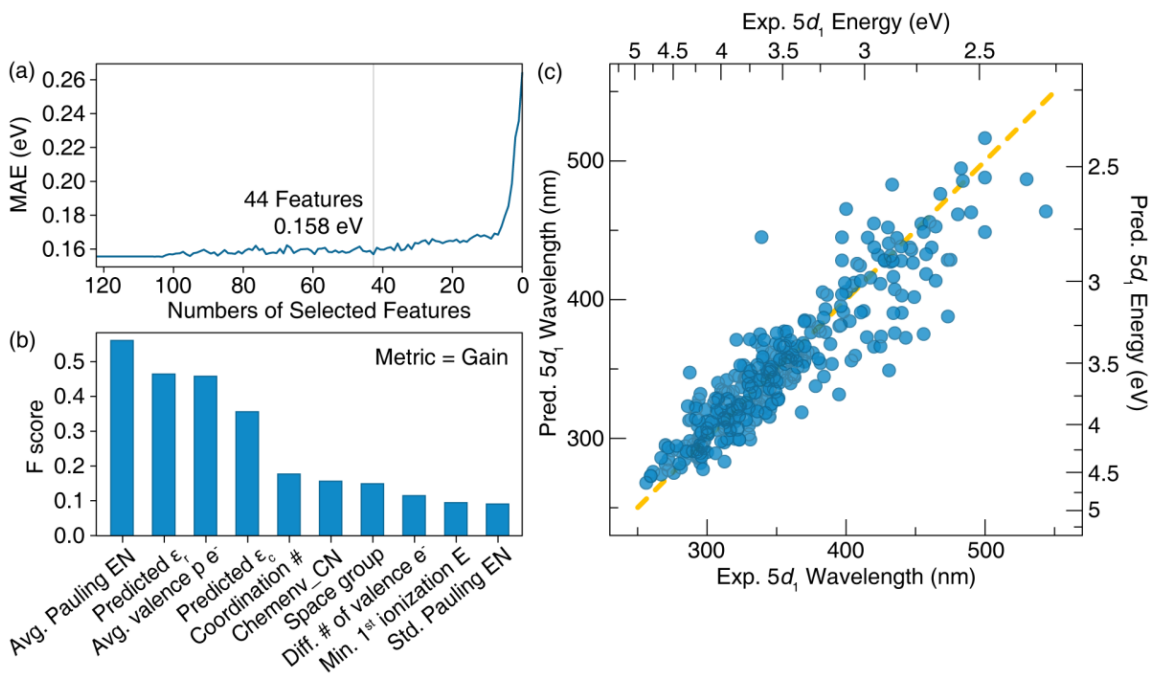

**Figure 2.** (a) RFE result with the initial XGBoost model. 44 features were chosen with minimal MAE. (b) Top 10 important features of this prediction model in terms of gain (F score). (c) Plot of experimental $5d_1$ energy versus

predicted $5d_1$ by the final model. The MAE of this initial model is ±0.153 eV with $R^2$ of 83.8 %

A notable advantage of XGBoost, and machine learning in general, is the ability to analyze feature importance, thereby facilitating the interpretation of the relationship between targets and features. Therefore, a revised model was prepared using recursive feature elimination to iteratively remove the least important feature(s) and track model performance with each refined feature set. **Figure 2a** illustrates the change in model performance starting from the complete 124 feature set. Removing the least important feature(s) in each step shows that the model performance remains nearly constant based on the MAE until there are only ~40 features. Further feature reduction causes the model's error to exponentially increase when there are fewer than 10 features. A balance of features to performance suggests that 44 features are a reasonable number to build a reliable model. Analyzing the feature importance scores for the top 10 out of these 44 features indicates that the most critical features include the composition's average Pauling electronegativity, reflecting the host structure's nephelauxetic effect (**Figure 2b**). This is followed in importance by the predicted relative permittivity and centroid shift, corroborating the direct influence of the nephelauxetic effect on the $5d_1$ position. The fifth most important feature, coordination number, supports the influence of crystal field splitting on the $5d_1$ energy level. The next feature, chemenv_CN, also relates to the coordination environment but considers geometrical properties too.[55] Table S1 also provides all 44 selected features, marked with bold text. The physical relevance of these significant features to the $5d_1$ energy level supports and enhances our understanding of the underlying physical mechanisms governing the $4f\rightarrow5d_1$ transition.

The final operational XGBoost model re-trained using the 44 selected features exhibited an MAE of ±0.153 eV (RMSE = 0.154 eV) and an $R^2$ of 83.8 %, indicating satisfactory model performance (**Figure 2c**). This model shows modest improvements, but the smaller feature set reduces the likelihood of overfitting. The remaining prediction uncertainty likely stems from various sources. In some respects, it is simply a function of the limited data set size. However, it could also have originated from the raw experimental training data used to build the model. During the data cleaning process that was conducted before building the operational model, several training data problem cases were identified that could generate uncertainty. First, significant discrepancies were observed in the reported $5d_1$ values when comparing the same compound across multiple independent publications and data collected in our lab during reproduction efforts. Sometimes the most reliable data was found in archived literature, while in-house measurements were more reliable in other instances. Secondly, publications rarely provide the excitation spectrum across different $Ce^{3+}$ concentrations, which can affect the excitation peak position by altering lattice parameters or preferred site selectivity. For example, the excitation spectrum of $NaCaBO_3:Ce^{3+}$ exhibited spectral broadening across six different $Ce^{3+}$ concentrations (0.5 % to 5%), with $\lambda_{ex,max}$ shifting between 347 nm to 355 nm,[56] whereas $BaSi_7N_{10}:Ce^{3+}$ provided a $\lambda_{ex,max}$ of 325 nm at 0.5% and 1% $Ce^{3+}$ concentration, respectively.[57] However, many studies only provided one excitation spectrum at one optimal $Ce^{3+}$ concentration. Consequently, the machine learning model did not employ the $Ce^{3+}$ concentration as a feature, even though this information would likely improve the model's predictive accuracy. Furthermore, the $4f\rightarrow5d_1$ excitation peak is not always clearly reported in the literature or there can be conflicting reports. For instance, one publication reported a $\lambda_{ex,max}$=320 nm for $CaAl_2B_2O_7:Ce^{3+}$.[58] In contrast, another paper suggested that the strongest excitation peak can be deconvoluted into a strong peak at 310 nm ($5d_2$) and a weak peak at 338 nm ($5d_1$).[59] Data that were were not abundantly unambiguous were removed from the training data set. Observations such as these show the complexity of accurately building a prediction model through experimental data and highlight the need for careful consideration of data selection in future refinements of the predictive model.

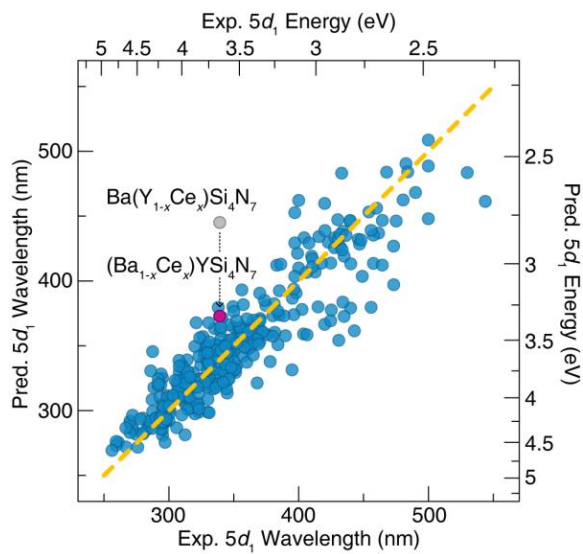

**Figure 3.** Plot of experimental $5d_1$ energy versus predicted $5d_1$ by the final model with fixed cation site information for $BaYSi_4N_7:Ce^{3+}$. This model with a hypothetical "$(Ba_{1-x}Ce_x)YSi_4N_7$" has an MAE of ±0.153 eV and an $R^2$ of 83.8 %

The model serves as more than just a regression prediction tool. It can also provide critical insights by evaluating reported data. For instance, during the first training phase, the data set contained $CaSnO_3:Ce^{3+}$ with an excitation peak position of 246 nm (5.04 eV), as reported in the literature.[60] During early training runs, this compound was one of the most significant outliers in our trained model. Reproducing this compound in-house revealed the experimental excitation peak was closer to 378 nm (3.28 eV), which makes more crystal-chemical sense. Subsequent training iterations incorporated this updated value, and the model's predictions aligned with experimental results, underscoring the use of data-driven techniques to refine and enhance data reliability.

Additionally, this model can aid in analyzing experimentally measured excitation spectra to help determine the most likely substitution site of an activator ion for phosphors containing multiple possible cation substitution sites. In systems with no strong energetic preference for the activator ion to substitute on a specific crystallographic site, the lanthanide may occupy multiple crystallographic positions concurrently, leading to an ensemble optical response. This phenomenon is evident in the training dataset, which includes 13 compounds reported with $Ce^{3+}$ on multiple unique crystallographic substitution sites. Determining the exact location of lanthanide ions in the host crystal structure to deduce their coordination environment requires various complementary spectroscopic techniques. Yet, sometimes, the results may not be conclusive, and a best guess is provided. An interesting outcome following this type of substitutional assignment was observed with this model for $BaYSi_4N_7:Ce^{3+}$, shown as the highlighted point in **Figure 3**.[61] Conventional phosphor design principles suggest that the $Ce^{3+}$ should preferentially substitute for cation sites with similar size, oxidation state, and coordination number. Some evidence is provided in the original report[61] supporting that $Ce^{3+}$ may occupy the $Y^{3+}$ site; however, when building the model with $Ce^{3+}$ on the $Y^{3+}$ site, a substantial discrepancy of 0.87 eV between the experimental (339 nm; 3.66 eV) and predicted (445 nm; 2.79 eV) $5d_1$ values was found, rendering it the most significant outlier. A hypothetical model was created here that placed $Ce^{3+}$ on the $Ba^{2+}$ site instead. Even though they have different oxidation states, $Ce^{3+}$ substituting for $Ba^{2+}$ is relatively common, and this hypothetical model yielded an expected value of 373 nm (3.32 eV), closely matching the experimental measurement. This finding demonstrates that machine learning can aid in interpreting optical properties and offer valuable insights into experimental results.

## 3.3 Screening for $Ce^{3+}$ Substituted Inorganic Phosphors with Target Excitation

To demonstrate the robustness, reliability, and utility of the final model for phosphor identification, the model was used to screen a wide dataset of candidate materials prior to experimentally investigating a promising candidate. All 153,188 materials listed in the Materials Project database,[51] including both experimentally reported and computationally predicted structures, were initially considered and then refined according to a set of heuristic design rules. The first of these rules was based on host composition. Potential phosphors were required to include the yellow-marked elements in **Figure 4**, which represent likely $Ce^{3+}$ substitution sites based on the literature.[12] These elements all have sufficiently large ionic radii and reasonable coordination numbers ranging from $r_{CN=6}$=1.01 Å to $r_{CN=12}$=1.34 Å to support $Ce^{3+}$ substitution.[62] The polyhedral backbone of the host was then limited to elements marked in light blue (cations) and dark blue (anions) based on common host chemistries contained in the training data. $Y^{3+}$ and $Lu^{3+}$ have been reported as both substitution sites or backbone cations due to their slightly smaller ionic radii than $Ce^{3+}$. This list also excludes elements that could interfere with $Ce^{3+}$ photoluminescence (*e.g.*, $Cr^{3+}$, $Mn^{2+}$, $Fe^{3+}$),[63–65] elements like Pb, Tl, and Cd due to sustainability and toxicity concerns, and expensive metals like Au, Pt, and Ir. After this filtering, 21,682 compounds that included at least one yellow and one dark blue-marked element, with optional light blue elements, were selected.

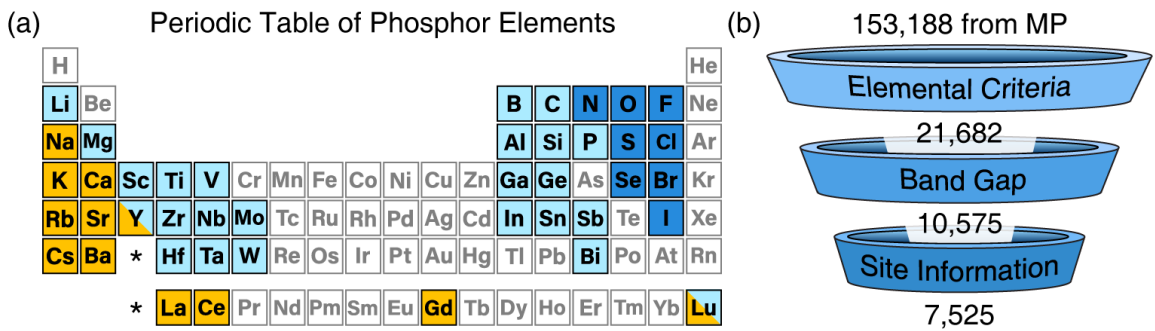

**Figure 4.** (a) The target elements for candidate phosphor host compositions. Elements forming the polyhedral backbone of the host structure are marked with the cation as light blue and the anion as dark blue. The possible $Ce^{3+}$ substitution site cations are marked in yellow. (b) Schematic diagram of selecting possible host structures from Materials Project.

In addition to elemental criteria, further screening was applied to ensure the suitability of the host compounds for $Ce^{3+}$ photoluminescence. Specifically, to prevent (thermal) ionization-induced temperature-dependent quenching of the $Ce^{3+}$ emission, the bandgap ($E_g$) should be larger than the energy of target excitation wavelengths. Given that this work targets 450 nm excitation (2.76 eV), only host structures with calculated electronic $E_{g,DFT}$ exceeding 2.1 eV were retained. This threshold accounts for the well-known underestimation of band gaps at the PBE level (which are available in the Materials Project). Considering this condition, the candidate set was further reduced to 10,575 compounds, ensuring a more reliable selection of potential hosts.[66] Next, compounds present in the training set or exhibiting improper crystal chemistry (*e.g.*, abnormal coordination numbers not purged in earlier down selection) were removed, reducing the dataset to 7,525 possible host crystal structures. Of these candidate hosts, around 90% have multiple crystallographically-distinct likely substitution sites for $Ce^{3+}$, expanding the final set of potential lanthanide substitutions to 54,885 possible locations for $Ce^{3+}$, for which their $5d_1$ energy was predicted.

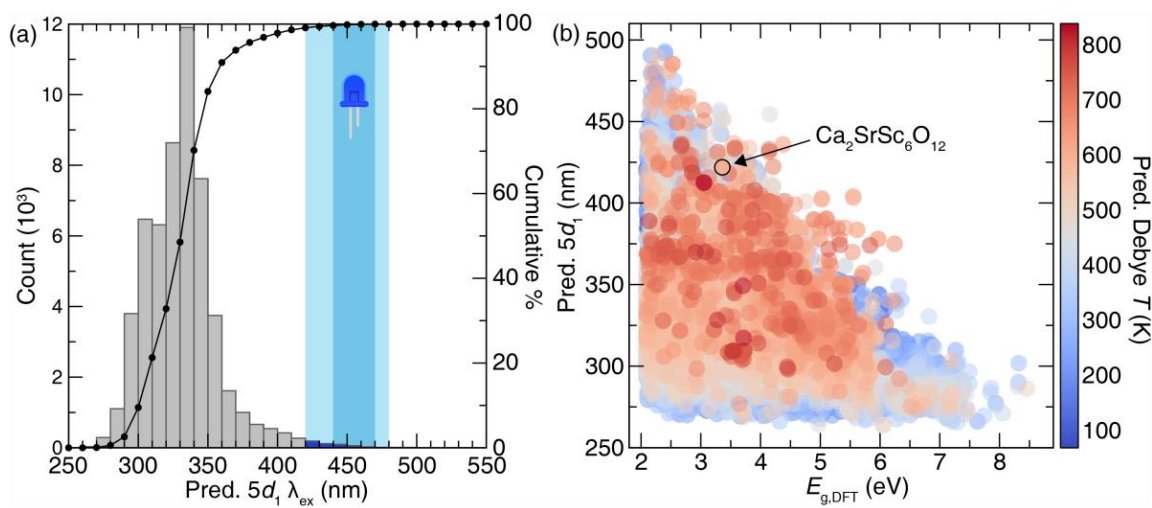

**Figure 5.** (a) Distribution of predicted excitation wavelengths for the 54,885 $Ce^{3+}$-substituted cation sites of possible host structures. The blue-shaded region highlights the InGaN LED emission range. (b) A plot mapping the DFT-calculated band gap (PBE-level), predicted Debye temperature, and the $5d_1$ prediction. The black-edged data point demonstrates $Ca_2SrSc_6O_{12}$:$Ce^{3+}$.

Plotting these 54,885 potential $Ce^{3+}$-substituted cation sites from the 7,525 unique host candidates in **Figure 5a** reveals that most $Ce^{3+}$ phosphors are excited in the UV range, agreeing with the literature. Only 494 cation sites (~0.9%) in 211 unique host structures (2.8%) were predicted to have blue excitation, highlighting the challenge of discovering blue-excited $Ce^{3+}$-based phosphors. Plotting the predicted excitation wavelengths against the host anion type showed that the most promising candidates are nitride/selenide/sulfide-containing compounds or garnet-based oxide crystal structures with gallium or germanium (**Figure S3**). These are classically understood to have the most significant redshifts and compatibility with blue LEDs.[8,21,24]

Selecting a blue-excited phosphor from this carefully screened list appears straightforward once the machine-learning predictions are made. However, compatibility with a blue LED chip doesn't guarantee practicality. To further accelerate phosphor host selection, this model was paired with additional proxies for phosphor performance as part of our selection pipeline. **Figure 5b** presents a sorting diagram of predicted $5d_1$ values versus the host material's band gap calculated at the DFT(PBE) level of theory ($E_{g,DFT}$). A wide bandgap is crucial for thermally robust $Ce^{3+}$ photoluminescence. An interesting trend is observed here. There is a negative correlation between excitation wavelength and bandgap. $Ce^{3+}$ substituted compounds with very wide bandgaps (~6 eV) are unlikely to be useful with blue LED chips due to high ionicity, which leads to a small centroid shift and weak crystal field splitting. Therefore, finding materials with sufficiently long wavelength excitation for industry applications will require a careful balance with the host material's bandgap. A second proxy that is well-regarded for correlating with a phosphor's photoluminescent quantum yield (PLQY) is the host structure's Debye temperature.[10,12] Phosphors with relatively high Debye temperatures and, thus, greater structural rigidity tend to exhibit higher PLQYs. Our group previously developed a machine learning model to predict the Debye temperature of host structures.[5] By incorporating this model into our analysis, as shown by the color of the data points in **Figure 5b**, we observe that many materials within the blue LED excitation range are predicted to have moderate to low Debye temperatures (≤ 500 K). However, some data points exhibit Debye temperatures greater than 600 K, indicating the potential to discover thermally robust, high photoluminescence quantum yield (PLQY) novel phosphors within this prediction set. Upon further analysis, many of the materials in this range are associated with garnet-type crystal structures or are nitrides, which are common blue-excited $Ce^{3+}$ phosphor hosts. The list of predicted $E_{g,DFT}$, Debye $T$, and $5d_1$ levels is available on GitHub (https://github.com/BrgochGroup).

After a thorough review of potential phosphor host compounds from this screened set, a novel composition was identified for synthesis. $Ca_2SrSc_6O_{12}$:$Ce^{3+}$ stood out due to its unique structure compared to other candidates while still suggesting impressive performance potential ($5d_1$ = 2.94 eV or 422 nm, an $E_{g,DFT}$ = 3.39 eV, and a Debye temperature = 594 K).

### 3.4 Experimental Validation of the $5d_1$ Model

The experimental validation of the selected potential inorganic phosphor involved solid-state synthesis. $Ca_2SrSc_6O_{12}$ crystallizes in the orthorhombic space group *Pnma* (no. 62) with a single crystallographically distinct statistically-mixed [(Ca/Sr)$O_8$] substitution site (**Figure 6a**).[67] Synchrotron powder X-ray diffraction was collected for $Ca_2SrSc_6O_{12}$:$Ce^{3+}$ and analyzed using the Rietveld refinement method starting from the published $Ca_2SrSc_6O_{12}$ structure file (PCD #1934108). $Ce^{3+}$ was omitted from the refinement due to its low concentration. The diffractogram verified the sample's purity (**Figure 6b**), while the refined lattice parameters provided in **Tables S2** and **S3** agree with the original structure. The $5d_1$ energy prediction was performed using the experimentally refined structure, accounting for site disorder with pymatgen. This yielded a prediction of 422 nm (2.94 eV) for $Ce^{3+}$ substitution on both the [$CaO_8$] and [$SrO_8$] sites, closely matching the original prediction. Experimentally measuring the photoluminescence excitation spectrum showed the peak falling at a maximum ($\lambda_{max}$) of 440 nm (2.82 eV), deviating by only 18 nm (0.12 eV) from the predicted value—well within the model's MAE (**Figure 6c**).

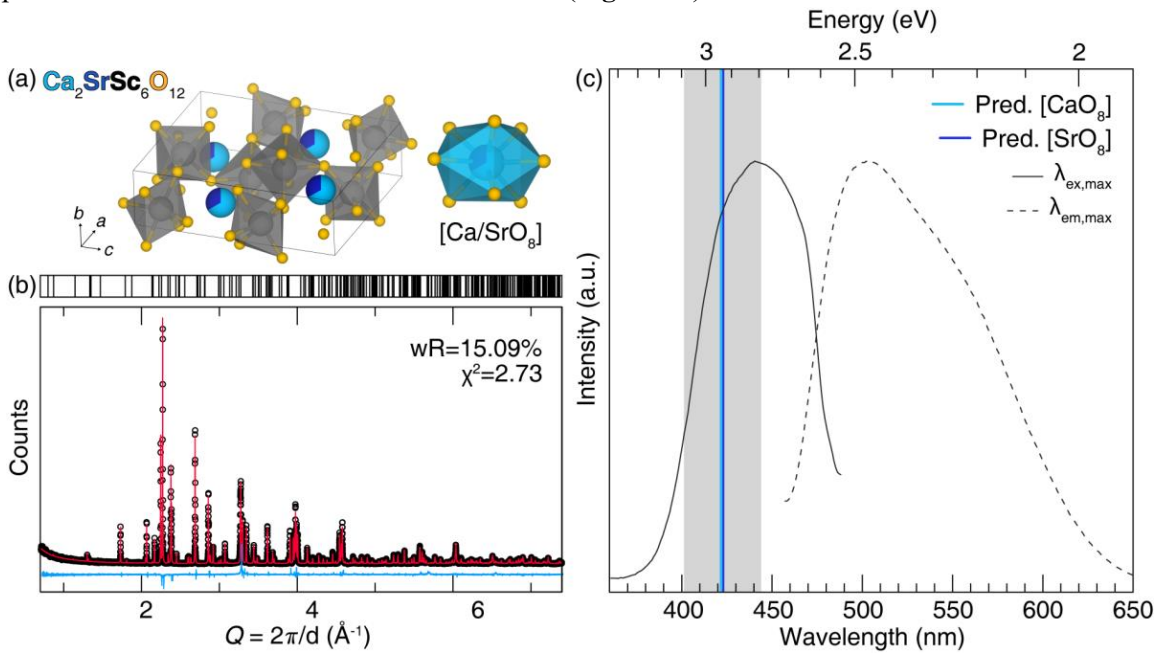

**Figure 6.** (a) Crystal structure of $Ca_2SrSc_6O_{12}$ (PCD #1934108) and $Ca^{2+}$/$Sr^{2+}$ site for the prediction. (b) Rietveld refinement of $Ca_2SrSc_6O_{12}$:$Ce^{3+}$ starting from PCD #1934108. (c) Normalized excitation and emission spectrum of $Ca_2SrSc_6O_{12}$:$Ce^{3+}$ with predicted $5d_1$ energy in wavelength scale of [$CaO_8$] and [$SrO_8$] site. The predicted $5d_1$ of [$CaO_8$] is 422 nm, while [$SrO_8$] is 422 nm. The experimental excitation peak position is 440 nm, and the emission peak is 503 nm.

The remaining optical properties were analyzed with a blue LED to assess the phosphor's practical performance. Collecting the room temperature photoluminescence emission ($\lambda_{ex}$ = 440 nm) spectrum revealed the phosphor exhibits a light green color ($\lambda_{em,max}$=503 nm) with an extremely broad (112 nm; 4,000 cm$^{-1}$) *fwhm*, spanning from 450 nm to 650 nm (**Figure 7a**). The emission spectrum at 100 K was deconvoluted into two distinct peaks ($\lambda_{max,1}$= 497 nm; 20,134 cm$^{-1}$, $\lambda_{max,2}$= 553 nm; 18,074 cm$^{-1}$) separated by ≈2,000 cm$^{-1}$ as expected for the spin-orbit-coupled $5d_1 \rightarrow {}^2F_{5/2}, {}^2F_{7/2}$ transitions of Ce$^{3+}$.[68,69] Measuring the temperature-dependent photoluminescence, plotted in **Figure 7b**, shows the two peaks merge due to thermal broadening as the temperature increases. The emission peak also shifted slightly from 503 nm at 100 K to 517 nm at 650 K. These changes lead to a shift in the CIE coordinates falling outside the 3-step MacAdam ellipse across the entire temperature range, indicating the change in emission color as a function of temperature is noticeable to the standard human observer (**Figure 7c**). Analyzing the change in emission intensity over the specified temperature range, as plotted in **Figure 7d**, reveals that the optical response of the phosphor was only slightly affected from 100 K to room temperature, with a decrease of approximately 5% before the thermal performance of the phosphor is impacted. The $T_{50}$, defined here as the temperature at which the luminescence intensity reaches 50% of its maximum value, was found to be 450 K, surpassing the current expectations set by the U.S. Department of Energy (423 K).[3]

The measured PLQY is slightly low at 33(2) %. However, it is essential to emphasize that this synthesis is unoptimized, as it lacks post-processing, fluxes/mineralizers, or other conventional methods commonly employed to enhance phosphor performance. The lower PLQY could also be attributed to charge-compensating defects resulting from Ce$^{3+}$ substituting for Ca$^{2+}$/Sr$^{2+}$. These defects are known to quench the PLQY, yet aliovalent substitutions are common in literature. To improve the performance of this phosphor, strategies such as co-substitution with charge-compensating ions or additional chemical modifications to minimize the detrimental defects and enhance the PLQY, ultimately optimizing the material for practical use. Finally, photoluminescence lifetime measurement was collected for completeness and yielded a decay time of 37.8(6) ns (**Figure S4**). Overall, the performance of Ca$_2$SrSc$_6$O$_{12}$:Ce$^{3+}$ aligns with the expected performance based on the calculated Debye temperature and (PBE-level) band gap,[12] and there is excellent agreement with the new $5d_1$ model created here, ensuring blue-LED compatibility.

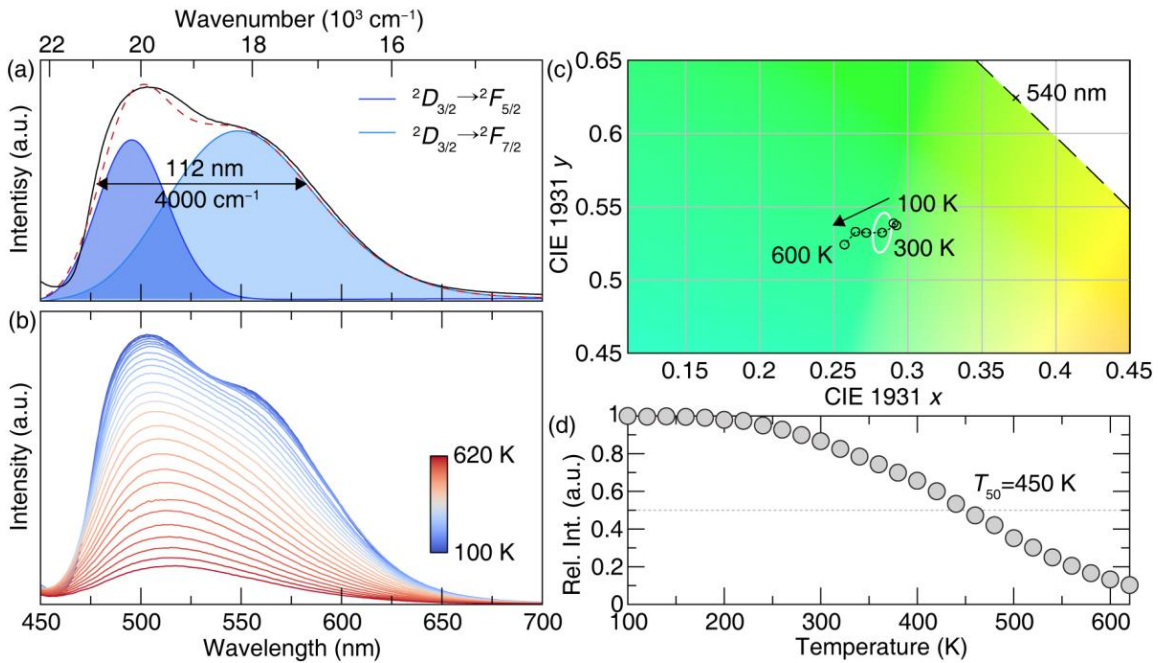

**Figure 7.** (a) Deconvoluted emission spectrum of $Ca_2SrSc_6O_{12}:Ce^{3+}$ at 100 K under 450 nm excitation. (b) Temperature-dependent emission spectra measured from 100 K to 620 K. (c) The emission spectra from 100 K to 600 K with 100 K interval, plotted in CIE 1931 space, showing chromatic stability of $Ca_2SrSc_6O_{12}:Ce^{3+}$. (d) Relative integrated emission intensity as a function of temperature.

## 4. Conclusion

This study introduced a supervised machine-learning approach for predicting the $5d_1$ excitation energy levels in $Ce^{3+}$ phosphors. The XGBoost model was trained on a diverse dataset comprising 357 experimentally measured $5d_1$ energy levels from 337 different host structures, achieving a mean prediction accuracy of ±0.15 eV. This high level of performance was accomplished through feature engineering, leave-one-group-out cross-validation techniques, and recursive feature elimination, resulting in a refined set of 44 scientifically relevant features. Feature importance analysis provided further insights into the physical mechanisms determining the $5d_1$ energy levels, including average Pauling electronegativity, relative permittivity, and centroid shifts, among others. Beyond its capabilities for linear regression prediction, this $5d_1$ model also demonstrated effectiveness as a tool for analyzing the excitation spectra of multi-cation site phosphors and for interpreting outlier detection. Finally, this new machine learning model allowed the rapid prediction of more than 50,000 possible phosphor $5d_1$ energies, leading to the successful identification, synthesis, and characterization of a novel, blue-excited phosphor, green-emitting composition: $Ca_2SrSc_6O_{12}:Ce^{3+}$. This discovery was the ultimate validation of the model's effectiveness in practical phosphor discovery.

*ORCID*
Nakyung Lee: 0000-0001-5048-1816
Małgorzata Sójka: 0000-0001-9346-8929
Seán Kavanagh: 0000-0003-4577-9647
Docheon Ahn: 0000-0003-1814-5418
David Scanlon: 0000-0001-9174-8601
Jakoah Brgoch: 0000-0002-1406-1352

*Notes*
The authors declare no competing financial interest.

## 5. Acknowledgments

The authors thank the U.S. National Science Foundation (DMR-2349319), the Welch Foundation (E-2181), the Texas Center for Superconductivity at the University of Houston (TcSUH), and the National Research Foundation of Korea (NRF) grant funded by the Korean government (MSIT) (RS-2024-00419831) for supporting this work.

## 6. Data Availability
The $Ce^{3+}$ $5d_1$ raw training data, feature generation code, and prediction model GitHub (https://github.com/BrgochGroup/Ce_5d1_Prediction) along with a permanent version of record at Zenodo (https://doi.org/10.5281/zenodo.14872504).

## 7. Supporting Information

The Supporting Information is available free of charge at https://.

- Plots of the revised relative permittivity, centroid shift models, and the initial 5$d$1 prediction model before recursive feature elimination, list of feature names for the initial and after RFE feature sets, predicted value distribution by the host anion type, Rietveld refined results and lifetime measurement at room temperature for $Ca_2SrSc_6O_{12}$:$Ce^{3+}$.

# Supporting Information

# Machine Learning a Phosphor's Excitation Band Position

Nakyung Lee[1,2,⊥], Małgorzata Sójka[1,2,⊥], Annie La[1,2], Syna Sharma[1,2], Seán Kavanagh[3], Docheon Ahn[4], David O. Scanlon[5,*], Jakoah Brgoch[1,2,*]

[1]*Department of Chemistry, University of Houston, Houston, Texas 77204, USA*
[2]*Texas Center for Superconductivity, University of Houston, Houston, Texas 77204, USA*
[3]*Harvard University Center for the Environment, Cambridge, Massachusetts 02138, United States*
[4]*PLS-II Beamline Department, Pohang Accelerator Laboratory, POSTECH, Pohang 37673, Republic of Korea*
[5]*School of Chemistry, University of Birmingham, Birmingham, UK*

[⊥]These authors contributed equally to this work

*jbrgoch@uh.edu (J. Brgoch)

*d.o.scanlon@bham.ac.uk (D. O. Scanlon)

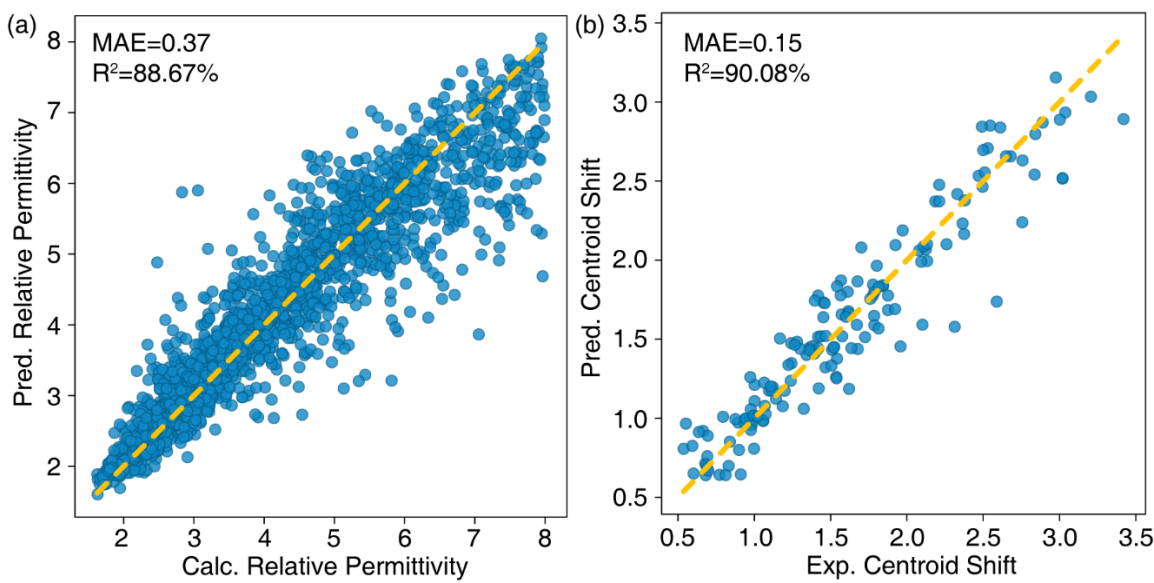

**Figure S1.** Leave-One-Out cross-validation of (a) the reconstructed relative permittivity prediction model and (b) the reconstructed centroid shift prediction model.

**Table S1.** Initial feature set for the $5d_1$ prediction model. The final 44 features via RFE for the final model marked with bold font.

| Index | Feature Name | Index | Feature Name |
| --- | --- | --- | --- |
| **0** | **Predicted centroid shift** | 24 | **b** |
| **1** | **Predicted relative permittivity** | 25 | c |
| 2 | Polyhedron volume | 26 | alpha |
| **3** | **Coordination number** | 27 | beta |
| 4 | Point group | **28** | **gamma** |
| **5** | **Cation ionic radius** | 29 | Avg. atomic weight |
| **6** | **Dopant ionic radius** | **30** | **Avg. Mendeleev number** |
| 7 | Ionic radii difference | 31 | Avg. crystal radius |
| 8 | Min. metal ligand bond length | **32** | **Avg. Pauling electronegativity** |
| 9 | Max. metal ligand bond length | **33** | **Avg. number of valence electrons** |
| 10 | Mean. Metal ligand bond length | **34** | **Avg. Gilmor number of valence electron** |
| 11 | Distortion index | 35 | Avg. valence s |
| 12 | Cation site Madelung potential | **36** | **Avg. valence p** |
| 13 | CSM | **37** | **Avg. valence d** |
| **14** | **Chemenv_CN** | **38** | **Avg. 1st ionization potential (kJ/mol)** |
| **15** | **Space group number** | **39** | **Avg. polarizability (Å³)** |
| 16 | Crystal system | 40 | Avg. electron affinity (kJ/mol) |
| 17 | Polar axis | **41** | **Avg. density of element (g/ml)** |
| 18 | Inversion center | **42** | **Avg. specific heat (J/g·K)** |
| **19** | **Unit cell volume** | 43 | Avg. heat of fusion (kJ/mol) |
| **20** | **Unit cell volume per Z** | **44** | **Avg. heat of vaporization (kJ/mol)** |
| 21 | Unit cell volume per atom | 45 | Avg. thermal conductivity (W/m·K) |
| 22 | Density | 46 | Avg. heat atomization |
| **23** | **a** | 47 | Avg. cohesive energy |

| Index | Feature Name | Index | Feature Name |
|---|---|---|---|
| 48 | Diff. atomic weight | 86 | Min. atomic weight |
| 49 | Diff. Mendeleev number | 87 | Min. Mendeleev number |
| 50 | Diff. crystal radius | 88 | Min. crystal radius |
| 51 | Diff. Pauling electronegativity | 89 | Min. Pauling electronegativity |
| **52** | **Diff. number of valence electrons** | 90 | Min. number of valence electrons |
| **53** | **Diff. Gilmor number of valence electron** | 91 | Min. Gilmor number of valence electron |
| 54 | Diff. valence s | 92 | Min. valence s |
| 55 | Diff. valence p | 93 | Min. valence p |
| 56 | Diff. valence d | 94 | Min. valence d |
| 57 | Diff. 1$^{st}$ ionization potential (kJ/mol) | **95** | **Min. 1$^{st}$ ionization potential (kJ/mol)** |
| **58** | **Diff. polarizability (Å$^3$)** | 96 | Min. polarizability (Å$^3$) |
| **59** | **Diff. electron affinity (kJ/mol)** | 97 | Min. electron affinity (kJ/mol) |
| 60 | Diff. density of element (g/ml) | 98 | Min. density of element (g/ml) |
| **61** | **Diff. specific heat (J/g·K)** | 99 | Min. specific heat (J/g·K) |
| 62 | Diff. heat of fusion (kJ/mol) | 100 | Min. heat of fusion (kJ/mol) |
| **63** | **Diff. heat of vaporization (kJ/mol)** | **101** | **Min. heat of vaporization (kJ/mol)** |
| 64 | Diff. thermal conductivity (W/m·K) | 102 | Min. thermal conductivity (W/m·K) |
| 65 | Diff. heat atomization | **103** | **Min. heat atomization** |
| 66 | Diff. cohesive energy | 104 | Min. cohesive energy |
| 67 | Max. atomic weight | **105** | **Std. atomic weight** |
| 68 | Max. Mendeleev number | 106 | Std. Mendeleev number |
| **69** | **Max. crystal radius** | **107** | **Std. crystal radius** |
| 70 | Max. Pauling electronegativity | **108** | **Std. Pauling electronegativity** |
| 71 | Max. number of valence electrons | **109** | **Std. number of valence electrons** |
| 72 | Max. Gilmor number of valence electron | 110 | Std. Gilmor number of valence electron |
| 73 | Max. valence s | 111 | Std. valence s |
| 74 | Max. valence p | **112** | **Std. valence p** |
| 75 | Max. valence d | **113** | **Std. valence d** |
| 76 | Max. 1$^{st}$ ionization potential (kJ/mol) | 114 | Std. 1$^{st}$ ionization potential (kJ/mol) |
| 77 | Max. polarizability (Å$^3$) | 115 | Std. polarizability (Å$^3$) |
| 78 | Max. electron affinity (kJ/mol) | **116** | **Std. electron affinity (kJ/mol)** |
| 79 | Max. density of element (g/ml) | 117 | Std. density of element (g/ml) |
| 80 | Max. specific heat (J/g·K) | **118** | **Std. specific heat (J/g·K)** |
| **81** | **Max. heat of fusion (kJ/mol)** | 119 | Std. heat of fusion (kJ/mol) |
| **82** | **Max. heat of vaporization (kJ/mol)** | 120 | Std. heat of vaporization (kJ/mol) |
| 83 | Max. thermal conductivity (W/m·K) | 121 | Std. thermal conductivity (W/m·K) |
| 84 | Max. heat atomization | **122** | **Std. heat atomization** |
| 85 | Max. cohesive energy | 123 | Std. cohesive energy |

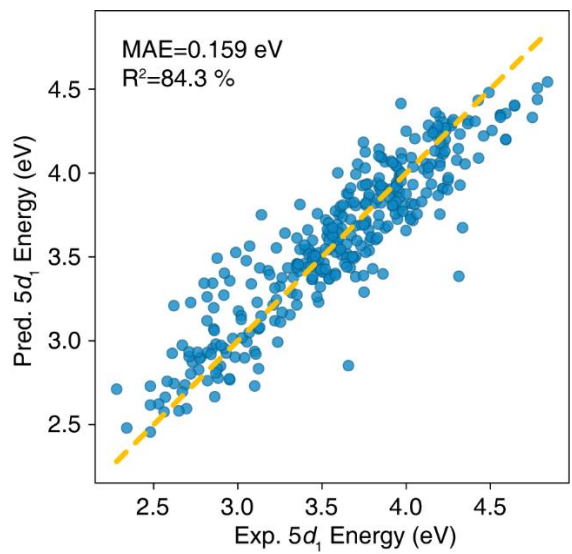

**Figure S2.** Plot of experimental $5d_1$ energy versus predicted $5d_1$ by the initial $5d_1$ prediction model with 124 features.

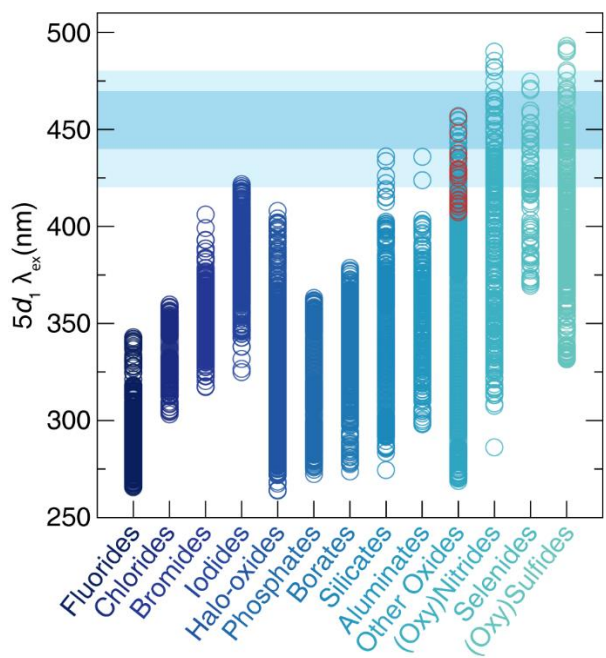

**Figure S3.** Distribution of predicted excitation wavelengths of $Ce^{3+}$ in the 54,885 cation sites screened from Materials Project, grouped by the host anion type. Red marked values are from the garnet structure type with gallium or germanium. The entire prediction set of excitation wavelength values is available on GitHub (https://github.com/BrgochGroup).

**Table S2.** Rietveld refinement data and statistics for $Ca_2SrSc_6O_{12}$

| Refined formula | $Ca_2SrSc_6O_{12}$ |
|---|---|
| Radiation λ (Å) | 1.546 |
| 2θ range (°) | 10-120 |
| Temperature (K) | 295 |
| Crystal system | Orthorhombic |
| Space group; Z | *Pnma*; 4 |
| a | 9.54741 |
| b | 3.16048 |
| c | 11.18849 |
| Volume (Å$^3$) | 337.606 |
| wR | 15.089 |
| $\chi^2$ | 2.73 |

**Table S3.** The refined atomic positions for Ca$_2$SrSc$_6$O$_{12}$

| Atom | Wyck. Pos. | x | y | z | U$_{iso}$ | Occ. |
|---|---|---|---|---|---|---|
| Ca1 | 4c | 0.2448(2) | 1/4 | 0.6535(1) | 0.0025(4) | 0.644(5) |
| Sr1 | 4c | 0.2448(2) | 1/4 | 0.6535(1) | 0.0025(4) | 0.356(5) |
| Sc1 | 4c | 0.0797(1) | 1/4 | 0.3939(1) | 0.0005(5) | 1.000 |
| Sc2 | 4c | 0.5705(2) | 1/4 | 0.6138(2) | 0.0013(5) | 1.000 |
| O1 | 4c | 0.0786(6) | 1/4 | 0.0832(5) | 0.011(2) | 1.000 |
| O2 | 4c | 0.2899(6) | 1/4 | 0.3276(5) | 0.009(2) | 1.000 |
| O3 | 4c | 0.3738(5) | 1/4 | 0.0219(5) | 0.003(2) | 1.000 |
| O4 | 4c | 0.4746(7) | 1/4 | 0.7858(5) | 0.015(2) | 1.000 |

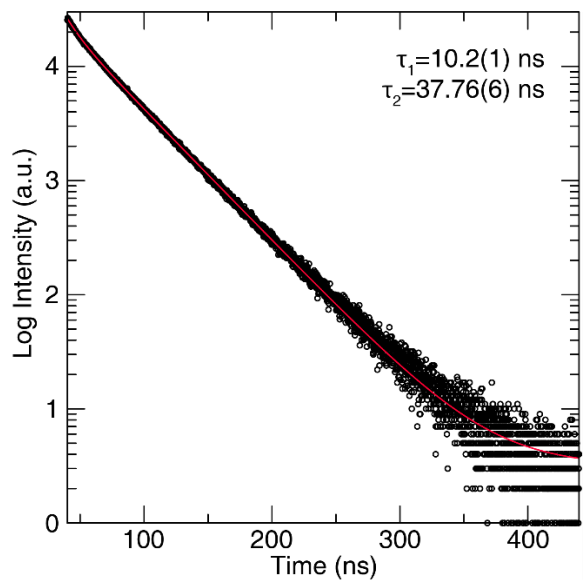

**Figure S4.** The lifetime measurement of Ca$_2$SrSc$_6$O$_{12}$:Ce$^{3+}$ at room temperature under 455 nm excitation, fitted with bi-exponential decay.